\documentclass[aps,prb,twocolumn,superscriptaddress,showpacs]{revtex4-1}

\usepackage{graphicx}
\usepackage{wrapfig}
\usepackage{amsmath}
\usepackage{amssymb}
\usepackage{bbm}
\usepackage{hyperref}

\begin{document}

\title{Magnetic response of energy levels of superconducting nanoparticles with spin-orbit scattering}

\author{Konstantin N. Nesterov}
\affiliation{Center for Theoretical Physics, Sloane Physics Laboratory, Yale University, New Haven, Connecticut 06520, USA}
\affiliation{Univ. Grenoble Alpes, INAC-SPSMS, F-38000 Grenoble, France}
\affiliation{CEA, INAC-SPSMS, F-38000 Grenoble, France}

\author{Y. Alhassid}
\affiliation{Center for Theoretical Physics, Sloane Physics Laboratory, Yale University, New Haven, Connecticut 06520, USA}

\date{\today}

\begin{abstract}
Discrete energy levels of ultrasmall metallic grains are extracted in single-electron-tunneling-spectroscopy experiments. We study the response of these energy levels  to an external magnetic field in the presence of both spin-orbit scattering and pairing correlations. In particular, we investigate $g$-factors and level curvatures that parametrize, respectively,  the linear and quadratic terms in the magnetic-field dependence of the many-particle energy levels of the grain.  Both of these quantities exhibit level-to-level fluctuations in the presence of spin-orbit scattering. We show that the distribution of $g$-factors is not affected by the pairing interaction and that the distribution of level curvatures is sensitive to pairing correlations even in the smallest grains in which the pairing gap is smaller than the mean single-particle level spacing. We propose the level curvature in a magnetic field as a tool to probe pairing correlations in tunneling spectroscopy experiments.
\end{abstract}


\maketitle

\newcommand{\hP}{{\hat{P}}}
\newcommand{\Tr}{{\mathrm{Tr}}}
\newcommand{\HBCS}{\hat{H}_{\mathrm{BCS}}}
\newcommand{\spinup}{{\uparrow}}
\newcommand{\spindown}{{\downarrow}}
\newcommand{\one}{{{\mathbbm{1}}}}
\newcommand{\sz}{{\hat{\sigma}_z}}

\section{Introduction}

Discrete electronic energy levels in individual nano-scale metallic grains (nanoparticles) can be measured by means of single-electron-tunneling spectroscopy.~\cite{vonDelft2001} Since the pioneering experiments in aluminum  grains,~\cite{Ralph1995,Black1996,Ralph1997}  spectra of nanoparticles made of a variety of materials were measured. In particular, the magnetic-field response of energy levels of grains with spin-orbit scattering has been studied in a number of experiments.~\cite{Salinas1999,Davidovic1999,Petta2001,Petta2002,Kuemmeth2008}

Spin-orbit scattering breaks spin symmetry.  However, time-reversal symmetry, leading to Kramers degeneracy of the single-electron levels in the absence of a magnetic field, remains a good symmetry. Since atomic-scale irregularities of a typical grain destroy all possible orbital symmetries, there are generally no additional degeneracies, and the single-electron levels form doublets.   An external magnetic field breaks time-reversal symmetry and leads to splitting of the doublets. The energies of the upward- ($\epsilon_{k+}$) and downward-moving ($\epsilon_{k-}$) levels of a doublet $\epsilon_k$
in a weak magnetic field $B$  are parametrized by the $g$-factor $g_k$ and level curvature $\kappa_k$ (at zero field) as
\begin{equation}\label{definition_sp}
 \epsilon_{k\pm}(B) = \epsilon_{k}(0) \pm \frac{1}{2} g_k\mu_B B + \frac{1}{2} \kappa_k B^2+ O(B^3),
\end{equation}
 where $\mu_B$ is the Bohr magneton.

 When spin-orbit coupling is negligible, the spin is a good quantum number, orbital magnetism does not contribute to $g$-factors,~\cite{Kravtsov1992, Matveev2000} and $g_k=2$ for all levels.
However, in the presence of spin-orbit coupling, the $g$-factor exhibits level-to-level fluctuations~\cite{Matveev2000,Brouwer2000,Adam2002} and depends on the magnetic-field direction.~\cite{Brouwer2000,Adam2002}
In addition, the spin contribution is suppressed and may become comparable to the contribution from orbital magnetism,~\cite{Matveev2000, Adam2002,Cehovin2004}  which is finite when spin symmetry is broken.
The statistics of single-particle $g$-factors have been studied using random-matrix theory (RMT),~\cite{Matveev2000,Brouwer2000,Adam2002} and are generally in good agreement with spectroscopy experiments on noble-metal nanoparticles.~\cite{Petta2001,Petta2002, Kuemmeth2008} However, the understanding of the average value of the $g$-factor is still lacking.

Another signature of spin-orbit scattering is the nonlinear second-order correction (in the magnetic field) to energies, which results in avoided crossing of energy levels.~\cite{Salinas1999, Adam2002} Although this correction is not identically zero in the absence of spin-orbit scattering because of orbital magnetism, it becomes appreciable only when spin-orbit scattering breaks spin-rotation symmetry, leading to finite matrix elements of the spin operator between states of different doublets. The distribution of level curvatures was measured in gold nanoparticles~\cite{Kuemmeth2008} and found to be in  agreement with RMT predictions.~\cite{Fyodorov1995, vonOppen1995, Mucciolo2006}

The above single-particle level results are valid for grains that are well described by the constant-interaction (CI) model, in which the electron-electron interaction is described by the classical charging energy ${e^2 {N}^2/2C}$, where ${N}$ is the electron number and $C$ is the capacitance of the grain.
In the CI model, the energies extracted in a tunneling-spectroscopy experiment reduce to the  single-electron energies, as is the case of noble-metal nanoparticles.~\cite{Petta2001,Petta2002,Kuemmeth2008}
However, when the interaction effects beyond the CI model are important (e.g., in superconducting or ferromagnetic materials), the extracted energies in the tunneling-spectroscopy experiments do not reduce to single-particle quantities.

Interaction effects in a chaotic or weakly disordered grain are described by the universal Hamiltonian~\cite{Kurland2000,Aleiner2002} in the limit of large  dimensionless Thouless conductance $g_{\text{Th}} = E_{\text{Th}}/\delta \gg 1$. Here $E_{\text{Th}}$ is the Thouless energy, determined by the time it takes for an electron to cross the grain, and $\delta$ is the mean spacing between single-particle Kramers doublets. The one-body part of the universal Hamiltonian follows RMT statistics,~\cite{Alhassid2000} while the interaction consists of universal terms that are consistent with the symmetries of the one-body Hamiltonian.~\cite{Kurland2000,Aleiner2002,Alhassid2005_prb} In the absence of spin-orbit scattering and orbital magnetic field, these terms are the charging energy (as in the CI model), pairing interaction, and ferromagnetic spin-exchange interaction.  The mesoscopic transport properties of a grain described by this universal Hamiltonian were studied in Ref.~\onlinecite{Schmidt2008}, and its 
thermodynamic observables  were calculated in Ref.~\onlinecite{Nesterov2013} using a Hubbard-Stratononvich~\cite{Hubbard1959,Stratonovich1957} decomposition for the pairing interaction and a spin projection method~\cite{Alhassid2003} for the exchange interaction. In the absence of pairing correlations (i..e, when only charging energy and spin-exchange correlations are present),  various observables can be calculated in closed form by using the Hubbard-Stratonovich transformation employed in Refs.~\onlinecite{Burmistrov2010, Burmistrov2012} or by the spin projection method. 

 In the presence of strong spin-orbit scattering,  spin-rotation symmetry is completely broken, and the exchange term is absent from the universal 
Hamiltonian.  The effect of the 
exchange interaction on the $g$-factor statistics in the crossover between weak and strong spin-orbit scattering was studied in Refs.~\onlinecite{Gorokhov2003,Gorokhov2004}. 
Large $g$-factors were recently observed in ferromagnetic cobalt nanoparticles.~\cite{Gartland2013}

Spin-orbit scattering preserves time-reversal symmetry and therefore does not suppress pairing correlations. 
In this work we explore the effect of an attractive pairing interaction on the magnetic-field response of many-particle energy levels of a grain in the presence of strong spin-orbit scattering, where the exchange interaction is negligible. 

We assume the one-bottleneck geometry~\cite{vonDelft2001,Gorokhov2003,Gorokhov2004} of the tunneling-spectroscopy experiments, for which the rate of tunneling onto the grain is much smaller than the rate of tunneling off the grain. In this limit, 
the current is determined by processes of tunneling onto the grain in its ground state. When the tunneling occurs onto a grain with even particle number $N_e$, the difference
$\Delta E_{\Omega,0} = E^{N_e+1}_\Omega-E^{N_e}_0$ between the energies of a twofold-degenerate level $\Omega$ for an odd electron number $N_e+1$ and the non-degenerate even ground state $E^{N_e}_0$  is measured. This energy difference splits in an external magnetic field. We define the many-body $g$-factor $g$ and level curvature $\kappa$ by generalizing the single-particle expression in (\ref{definition_sp})
\begin{equation}\label{definition}
\Delta E_{\Omega,0}(B) = \Delta E_{\Omega,0}(0) \pm \frac{1}{2}g\mu_B B + \frac{\kappa}{2}B^2 + O(B^3)\,.
\end{equation}
 
\begin{figure}[t]
\includegraphics[clip,width=0.45\textwidth]{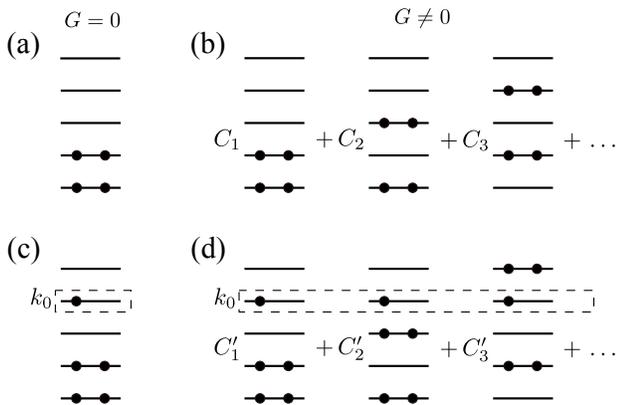}
\caption{Top: schematic description of the ground state of the Hamiltonian (\ref{BCS_Hamiltonian}) for an even number of particles in (a) the absence of pairing  ($G=0$) and (b)  the presence of pairing ($G\neq 0$). Bottom: a possible odd-particle-number state after the tunneling of an additional electron onto the even ground state in (c) the absence and (d) the presence of pairing. The solid lines describe doubly degenerate single-particle levels and the solid circles denote particles occupying these levels. }\label{Fig_states}
\end{figure}

In the absence of pairing correlations, the $N_e$-particle ground state consists of doubly occupied orbitals up to the Fermi level, while the $N_e+1$-particle level has one extra singly occupied orbital $k_0$ above the Fermi energy as is illustrated in panels (a) and (c) of Fig.~\ref{Fig_states}. In this limit, the $g$-factor and level curvature defined in Eq.~(\ref{definition}) reduce to the single-particle values $g_{k_0}$ and $\kappa_{k_0}$ of the singly occupied orbital $k_0$. In the presence of pairing correlations, the single-particle occupation numbers are no longer good quantum numbers. The even ground state becomes a superposition of Slater determinants describing fully paired noninteracting states as shown schematically in Fig.~\ref{Fig_states}(b). A final odd-particle-number state is a superposition of states of good occupation numbers with one singly occupied  ``blocked'' orbital $k_0$, which is the same in all the Slater determinants of the superposition~\cite{Soloviev1961} as shown in 
Fig.~\ref{Fig_states}(d).

 Here we will show that the $g$-factor as defined in (\ref{definition}) is identical to the single-particle $g$-factor $g_{k_0}$ of the blocked orbital $k_0$. On the other hand, the level curvature differs from its single-particle value, and its statistics are highly sensitive to pairing correlations. The main origin of this sensitivity is the change of the density of states induced by pairing correlations.

Another motivation of our studies is to identify signatures of pairing correlations in ultrasmall superconducting grains whose single-particle mean level spacing $\delta$ is comparable to or larger than the bulk pairing gap $\Delta$ of the grain. Anderson's criterion~\cite{Anderson1959_jpcs} states that superconductivity is no longer possible once $\Delta/\delta <1$, and the tunneling-spectroscopy experiments showed no traces of an excitation gap in the smallest aluminum grains that satisfy this condition.~\cite{Black1996} This criterion signifies the breakdown of the Bardeen-Cooper-Schrieffer (BCS) mean-field theory~\cite{Bardeen1957} in the fluctuation-dominated regime $\Delta/\delta \lesssim 1$. A better probe to detect pairing correlations in this regime are thermodynamic observables such as the heat capacity and spin susceptibility,~\cite{DiLorenzo2000,Falci2000,Nesterov2013} but they are difficult to measure experimentally. Here we propose the level curvature as a sensitive observable to probe pairing 
correlations in grains with spin-orbit scattering. Its advantage is that it can be measured directly in the tunneling-spectroscopy experiments.

The outline of the paper is as follows. In Sec.~\ref{Sec_model}, we introduce our model and discuss the many-body eigenstates of its corresponding Hamiltonian. In Sec.~\ref{Sec_gfactors}, we prove the robustness of $g$-factor statistics with respect to pairing correlations. In Sec.~\ref{Sec_curvatures}, we calculate the level curvature and discuss its statistics. We conclude in Sec.~\ref{Sec_conclusions}.

\section{Model}\label{Sec_model}

\subsection{Hamiltonian}

The universal Hamiltonian in the presence of strong spin-orbit scattering has the following form for fixed particle number
\begin{equation}\label{BCS_Hamiltonian}
\hat{H} = \sum_{k} \epsilon_k \left(c^\dagger_{k 1}c_{k 1} + c^\dagger_{k 2}c_{k 2}\right) - G\hP^\dagger \hP\;,
\end{equation}
where
\begin{equation}
\hP^\dagger = \sum_k c^\dagger_{k 1}c^\dagger_{k 2}\,\quad\text{and}\quad \hP = \sum_k c_{k 2}c_{k 1}\;
\end{equation}
are pair creation and annihilation operators. The single-particle states appear in degenerate time-reversed pairs $|k1\rangle_1$ and $|k2\rangle_1$ with single-particle energy $\epsilon_k$.  When spin symmetry is broken, they are no longer eigenstates of the spin-projection operator. We adopt the phase convention
\begin{equation}\label{phase_convention}
\hat{T}|k1\rangle_1 = |k2\rangle_1\,\quad \text{and} \quad \hat{T}|k2\rangle_1 = -|k1\rangle_1\;,
\end{equation}
where $\hat{T}$ is the time-reversal operator. The interaction term in Eq.~(\ref{BCS_Hamiltonian})  describes the scattering of electron pairs between such Kramers-degenerate orbitals and leads to BCS superconductivity in the bulk limit $\Delta/\delta \gg 1$. The exchange interaction is suppressed for strong spin-orbit scattering and is absent in Eq.~(\ref{BCS_Hamiltonian}).

The coupling of electrons in the grain to an external magnetic field along the $z$-axis is described by the Zeeman term
\begin{equation}
 \hat{V}(B) = -\hat{M}_zB\;,
\end{equation}
where $\hat{M}_z$ is the $z$-component of the magnetic-moment operator of those electrons. In general, the response of levels in the grain to an external magnetic field depends on its direction,~\cite{Brouwer2000,Adam2002} but we do not study this effect here.

In a chaotic or a weakly disordered grain, the single-particle levels follow RMT statistics. In the absence of an external magnetic field, time-reversal symmetry is preserved, and the single-particle spectrum is described by an ensemble that interpolates between the Gaussian orthogonal ensemble (GOE) and the Gaussian symplectic ensemble (GSE).~\cite{Mehta1991,Brouwer2000,Adam2002}  In the absence of spin-orbit scattering, spin-rotation symmetry is preserved and the corresponding ensemble is the GOE. In the presence of strong spin-orbit scattering, spin-rotation symmetry is completely broken and the corresponding ensemble is the GSE. We note, however, that  most of our qualitative conclusions are independent of the particular statistics of the single-particle eigenstates and eigenenergies.

In our quantitative calculations of level curvatures in Sec.~\ref{Sec_curvatures}, we make the following additional assumptions: (i) spin-orbit scattering is strong, i.e., the single-particle Hamiltonian follows the GSE statistics, (ii) the orbital contribution to the magnetic moment is negligible, i.e., $\hat{M}_z = 2\mu_B \hat{S}_z$, where $\hat{S}_z$ is the $z$-component of the spin operator, and (iii) in a finite-size model space with $N_{\text{sp}}$ Kramers-degenerate orbitals, the pairing coupling constant is determined from the relation~\cite{Berger1998,Alhassid2007_arxiv}
\begin{equation}
 \frac{G}{\delta} = \frac{1}{\mathrm{arcsinh}\left(\frac{N_{\text{sp}}/2}{\Delta/\delta}\right)}\,.
\end{equation}

\subsection{Many-body eigenstates}

In the absence of pairing correlations, each many-particle eigenstate of the Hamiltonian (\ref{BCS_Hamiltonian}) is a state with well-defined occupation numbers of single-particle orbitals. We will denote such noninteracting state by
$|\vec{m},k_1\alpha_1 \ldots k_t\alpha_t\rangle_N$, where $\vec{m} = \{m_1,\ldots,m_r\}$ is a set of $r$ doubly occupied orbitals,  $k_1,\ldots,k_t$ are $t$ singly occupied orbitals ($k_i \notin \vec{m}$) with corresponding labels $\alpha_1,\ldots,\alpha_t$ ($\alpha_i=1,2$) distinguishing Kramers-degenerate single-particle states, and $N =2r +t$ is the total number of particles.  

In the presence of pairing correlations, each eigenstate is a superposition of these noninteracting states. Since pair scattering cannot affect the singly occupied orbitals, the orbitals $k_1,\alpha_1,\ldots,k_t,\alpha_t$ are good quantum numbers and are the same in all the noninteracting states comprising the superposition of any given eigenstate of the Hamiltonian (\ref{BCS_Hamiltonian}). This is a manifestation of the blocking effect of the pairing interaction.~\cite{Soloviev1961} An $N$-particle eigenstate can thus be written as 
\begin{equation}\label{eigenstate_general}
 |\Psi, k_1\alpha_1 \ldots k_t\alpha_t\rangle_N = \sum_{\vec{m}: k_i \notin \vec{m}} C^\Psi_{\vec{m}} |\vec{m},k_1\alpha_1 \ldots k_t\alpha_t\rangle_N\,.
\end{equation}
  The sum in Eq.~(\ref{eigenstate_general}) runs over all sets $\vec{m}$ of $(N-t)/2$ doubly occupied orbitals that do not contain any of the singly occupied orbitals $k_i$. The coefficients $C^\Psi_{\vec{m}}$ can be chosen to be real. 
The energy of the state (\ref{eigenstate_general}) and the corresponding coefficients $C^\Psi_{\vec{m}}$ are independent of $\alpha_i$. Since each $\alpha_i$ can assume two values, the corresponding many-particle level is $2^t$-fold degenerate.

The ground state $\left|0\right\rangle_{N_e}$ of (\ref{BCS_Hamiltonian}) for an even number of particles $N_e$ is described by a superposition of fully paired noninteracting states with no singly occupied orbitals
\begin{equation}\label{eigenstate_even_ground}
\left|0\right\rangle_{N_e} = \sum_{\vec{m}} C^{0}_{\vec{m}} |\vec{m}\rangle_{N_e}\;,
\end{equation}
 and is schematically shown in Fig.~\ref{Fig_states}(b).
A possible odd state $\left|\Omega\right\rangle_{N_e+1}$ after tunneling onto the grain in $\left|0\right\rangle_{N_e}$ must have non-zero overlap with $c^\dagger_{k\alpha}\left|0\right\rangle_{N_e}$
 for at least one single-particle orbital $|k\alpha\rangle_1$. This, in combination with the blocking effect, dictates that such an  eigenstate has exactly one blocked (i.e., singly occupied) orbital as illustrated in Fig.~\ref{Fig_states}(d). These states form doublets and can be expressed as
\begin{equation}\label{odd_state}
|\Psi,{k_0\alpha_0}\rangle_{N_e+1}  = \sum_{\vec{m}: k_0\notin \vec{m}} C^{\Psi}_{\vec{m}}|\vec{m},k_0\alpha_0\rangle_{N_e+1}\,.
\end{equation}

\section{$g$-factor}\label{Sec_gfactors}

The many-particle  $g$-factor defined in Eq.~(\ref{definition})  depends, in general, on the linear corrections to both the odd- and even-particle energies. However, since the even state $|0\rangle_{N_e}$ is invariant under time reversal and $\hat M_z$ is odd,
\begin{equation}\label{Mz_even}
\left\langle 0 \left|\hat{M}_z\right|0\right\rangle_{N_e} = 0 \;,
\end{equation}
 and the linear correction to the even ground-state energy $E^{N_e}_0$ vanishes.
The $g$-factor that corresponds to the $\left|0\right\rangle_{N_e} \rightarrow \left|\Omega\right\rangle_{N_e+1} = \left|\Psi,{k_0\alpha}\right\rangle_{N_e+1}$ transition is then determined by the $2\times 2$ matrix of $\hat{M}_z$ written for the doublet $\left|\Psi, {k_0\alpha}\right\rangle_{N_e+1}$. From Eq.~(\ref{odd_state}), we find
\begin{multline}
\left\langle \Psi, {k_0\alpha} \left|\hat{M}_z\right| \Psi,{k_0\alpha'}\right\rangle_{N_e+1}  \\
= \sum_{\vec{m},\vec{m}': k_0\notin \vec{m},\vec{m}'}C^{\Psi}_{\vec{m}} C^{\Psi}_{\vec{m}'} \left\langle \vec{m},k_0\alpha \left|\hat{M}_z\right| \vec{m}',k_0\alpha'\right\rangle_{N_e+1}\,.
\end{multline}
Using 
\begin{equation}\label{one_body_operator}
 \hat{M}_z = \sum_{k\rho,k'\rho'} M^z_{k\rho,k'\rho'}c^\dagger_{k\rho} c_{k'\rho'}\,,
\end{equation}
 where $M^z_{k\rho,k'\rho'}$ are the single-particle matrix elements of $\hat M_z$, we have
\begin{multline}
\left\langle \vec{m},k_0\alpha \left|\hat{M}_z\right| \vec{m}',k_0\alpha'\right\rangle_{N_e+1}
= \delta_{\vec{m}\vec{m}'} \left[\vphantom{\sum} M^z_{k_0\alpha,k_0\alpha'} \right.\\ + \left. \delta_{\alpha\alpha'} \sum_{m\in\vec{m}} \left(M^z_{m1,m1}+M^z_{m2,m2}\right)\right]\,.
\end{multline}
Since the single-particle states $|m 1\rangle$ and $|m2\rangle$ are mapped onto each other under time reversal [see Eq.~(\ref{phase_convention})] and $\hat M_z$ is odd under time reversal,
\begin{equation}\label{Mz_sp_timereversal}
M^z_{m1,m1}+M^z_{m2,m2}=0\,.
\end{equation}
We conclude
\begin{equation}\label{Mz_elements}
\left\langle \Psi,k_0\alpha \left|\hat{M}_z\right| \Psi,k_0\alpha'\right\rangle_{N_e+1} = M^z_{k_0\alpha,k_0\alpha'}\,,
\end{equation}
where we have used the normalization condition $ \sum_{\vec{m}: k_0 \notin \vec{m}} \left(C^\Psi_{\vec{m}}\right)^2=1$.

 According to Eq.~(\ref{Mz_elements}), the matrix elements of the magnetic-moment operator between any two many-particle eigenstates that belong to the same Kramers doublet reduce to the single-particle matrix elements of the blocked orbital. Therefore, the  $g$-factor for the $|0\rangle_{N_e} \rightarrow |\Psi,k_0\alpha\rangle_{N_e+1}$ transition is exactly the single-particle $g$-factor of the blocked orbital $k_0$.

The above result is quite general. It is independent of the statistics of the single-particle levels, strength of spin-orbit scattering, or relative  weights of the spin and orbital parts in the magnetic moment. It follows from the blocking effect of the pairing interaction and from time-reversal symmetry. The blocking effect makes one orbital $k_0$ special in the odd state and separates the magnetic moment of that state into the contributions from a single electron on the orbital $k_0$ and from the remaining fully-paired electrons. Time-reversal symmetry makes the contributions from the paired electrons to the magnetic moments of the odd and even states zero.

Our result can be verified in a tunneling-spectroscopy experiment by measuring $g$-factors at different particle numbers (achieved by varying the gate voltage). Since the $g$-factor is completely determined by the blocked orbital $k_0$ of the final state of a tunneling process $|0\rangle_{N_e}\rightarrow |\Psi,k_0\alpha\rangle_{N_e+1}$, it remains the same for any other allowed process $|0\rangle_{\tilde{N}_e} \rightarrow |\tilde{\Psi},k_0\alpha\rangle_{\tilde{N}_e+1}$ with $\tilde{N}_e\ne N_e$ as long as the blocked orbital of the final state is the same.

Another consequence of our result is that the measured distribution of the $g$-factor reduces to the distribution of the single-particle $g$-factor. For sufficiently strong spin-orbit scattering,
the latter distribution has a universal form~\cite{Matveev2000,Brouwer2000} that should not be affected by pairing correlations. We note that the RMT predicts the distribution of $g$-factors measured in units of their average values $\langle g\rangle$, but not the average values themselves.

 A more general pairing interaction with orbital-dependent coupling constants of the form $\sum_{k,k'} G_{kk'}c^\dagger_{k1}c^\dagger_{k2}c_{k'2}c_{k'1}$ leads to blocking effects as well and does not modify the $g$-factor. Therefore, the $g$-factor can be used to probe electron-electron correlations beyond any such generalized pairing model.

In the absence of spin-orbit scattering, the exchange interaction $-J_s{\bf \hat{S}}^2$ (where ${\bf \hat{S}}$ is the total spin of the universal Hamiltonian) is consistent with the symmetries of the single-particle Hamiltonian and does not destroy the blocking effect. Finite spin-orbit scattering makes the structure of this term complex when written in the basis diagonalizing single-particle Hamiltonian [i.e., the $|k\alpha\rangle_1$ basis in Eq.~(\ref{BCS_Hamiltonian})] and destroys the blocking effect. The many-body eigenstates become superpositions of noninteracting states with different singly occupied orbitals, so the magnetic moment of such a state can no longer be reduced to a single-particle magnetic moment. The $g$-factors and their distribution are affected by the exchange interaction, as was studied in detail in  Refs.~\onlinecite{Gorokhov2003,Gorokhov2004}.

\section{Level curvature}\label{Sec_curvatures}

 Using second-order perturbation theory, the level curvature for the transition $|0\rangle_{N_e} \rightarrow |\Omega\rangle_{N_e+1}$ is given by
\begin{equation}\label{curvature_mb}
\kappa = \kappa_{\Omega}^{N_e+1} - \kappa_0^{N_e}\,,
\end{equation}
where
\begin{equation}\label{curvature_odd}
\kappa_{\Omega}^{N_e+1} = 2 \sideset{}{'}\sum_{\Omega'} \frac{\left|\left\langle \Omega\left|\hat{M}_z\right|\Omega'\right\rangle_{N_e+1}\right|^2}{E^{N_e+1}_{\Omega}-E^{N_e+1}_{\Omega'}}
\end{equation}
and
\begin{equation}\label{curvature_even}
\kappa_0^{N_e} =   2\sideset{}{'}\sum_{\Theta'} \frac{\left|\left\langle 0\left|\hat{M}_z \right|\Theta'\right\rangle_{N_e}\right|^2}{E^{N_e}_{0}-E^{N_e}_{\Theta'}}\,.
\end{equation}
Here the sums run over the many-electron eigenstates with energies different from $E^{N_e+1}_{\Omega}$ or $E^{N_e}_{0}$. The expressions (\ref{curvature_odd}) and (\ref{curvature_even}) can be thought of as the curvatures of the odd and even states, respectively.

In the $G\rightarrow 0$ limit, the curvature (\ref{curvature_mb}) reduces  to the single-particle curvature
\begin{equation}\label{curvature_sp}
\kappa_{k_0} = 2 \sum_{k\ne k_0} \frac{|M^z_{k_01,k1}|^2+|M^z_{k_01,k2}|^2}{\epsilon_{k_0} - \epsilon_k}\,,
\end{equation}
where $k_0$ is the blocked orbital in the final state $|\Omega\rangle_{N_e+1}$. 
The corresponding distribution $P(\kappa_{k_0})$ is symmetric around zero and was calculated analytically in the GSE limit.~\cite{Fyodorov1995,vonOppen1995}  The right (left) tail of $P(\kappa_{k_0})$ is determined by the probability that the energy of the orbital below (above)  $k_0$ is very close to $\epsilon_{k_0}$, which in the GSE limit results in $P(\kappa_{k_0}) \sim 1/\kappa_{k_0}^6$ in the tails of the distribution.~\cite{vonOppen1995}

In the presence of pairing, $P(\kappa)$ changes and depends on the final level $|\Omega\rangle_{N_e+1}$. When this level is the odd ground state $|0\rangle_{N_e+1}$, both contributions (\ref{curvature_odd}) and (\ref{curvature_even}) are negative. Therefore, $\kappa$ for the transition $|0\rangle_{N_e}\rightarrow |0\rangle_{N_e+1}$  is positive when $|\kappa_0^{N_e}| > |\kappa_0^{N_e+1}|$ and vice versa. Since pairing induces a gap in the even excitation spectrum only, $|\kappa_0^{N_e}|$ is suppressed by pairing, while $|\kappa_0^{N_e+1}|$ is not, making the distribution asymmetric with a negative median value. 
We verify this qualitative reasoning by calculating level curvatures exactly and in a generalized BCS approach for transitions to both the ground and excited states.

\subsection{Formalism}\label{Sec_curvatures_formalism}

Below we demonstrate the many-body formalism for the simple noninteracting case $G=0$, and then discuss the interacting case of finite $G$.  

\subsubsection{Noninteracting limit}

\begin{figure}[t]
 \includegraphics[clip,width=0.45\textwidth]{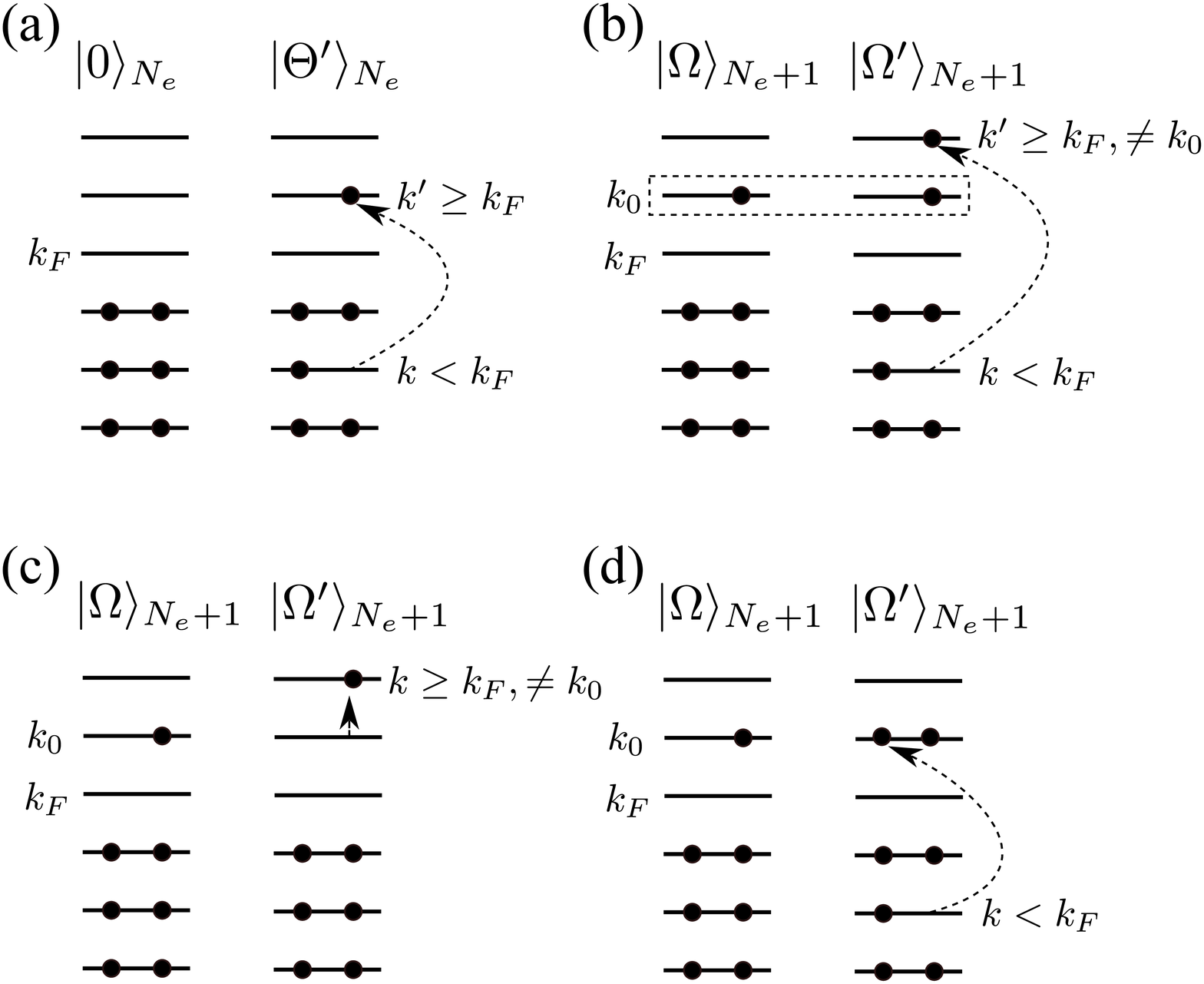}
 \caption{ Pairs of many-electron states contributing to the second-order perturbative expressions (\ref{curvature_odd}) and (\ref{curvature_even}) of the level curvature in the $G=0$ limit. For an even grain, the two states in a pair must differ by a particle-hole excitation that reduces by one the number of doubly occupied levels [panel (a)]. For an odd grain, these states may relate to each other as in (a) [see panel (b)] or they can have the same number of  doubly occupied levels but different blocked orbitals [panels (c) and (d)].}\label{Fig_matrixelements}
\end{figure}

Let $k_F$ be the lowest empty orbital in the noninteracting even ground state. 
Then, the states $|\Theta'\rangle_{N_e}$ contributing to Eq.~(\ref{curvature_even}) are particle-hole excitations $|\Theta'\rangle_{N_e} = c^\dagger_{k'\alpha'}c_{k\alpha}|0\rangle_{N_e}$ with $k'\ge k_F$ and $k<k_F$. A possible pair of states $|0\rangle_{N_e}$ and $|\Theta'\rangle_{N_e}$ is shown schematically in Fig.~\ref{Fig_matrixelements}(a). 
The corresponding matrix element between these states is $\langle \Theta'| \hat{M}_z|0\rangle_{N_e} = M^z_{k'\alpha',k\alpha}$, and their energy difference is $E^{N_e}_0 - E^{N_e}_{\Theta'}=\epsilon_k-\epsilon_{k'}$.
From time-reversal symmetry  $|M^z_{k'1,k2}| = |M^z_{k'2,k1}|$ and $|M^z_{k'2,k2}| = |M^z_{k'1,k1}|$ so the curvature of the even state is 
\begin{equation}\label{curvature_nonint_even}
 \kappa_0^{N_e}\left.\vphantom{\int}\right|_{G=0} = 4 \sum_{k<k_F\le k'}\frac{M_{kk'}}{\epsilon_{k}-\epsilon_{k'}}\,,
\end{equation}
where
\begin{equation}\label{M_kk'}
 M_{kk'}=|M^z_{k1,k'1}|^2+|M^z_{k1,k'2}|^2\,.
\end{equation}

There is a similar contribution to the curvature (\ref{curvature_odd}) of an odd state $|\Omega\rangle_{N_e+1}$ with a blocked orbital $k_0\ge k_F$ [see Fig.~\ref{Fig_matrixelements}(b)]
\begin{equation}\label{curvature_nonint_1}
 \kappa_{\Omega}^{(1)}\left.\vphantom{\int}\right|_{G=0} = 4 \sum_{k<k_F\le k', k'\ne k_0}\frac{M_{kk'}}{\epsilon_{k}-\epsilon_{k'}}\,.
\end{equation}
There is also a second contribution arising from  pairs of states that have different blocked orbitals but  the same number of doubly occupied levels [shown in Figs.~\ref{Fig_matrixelements}(c) and \ref{Fig_matrixelements}(d)]:
\begin{multline}\label{curvature_nonint_2}
  \kappa_{\Omega}^{(2)}\left.\vphantom{\int}\right|_{G=0} = 2\sum_{k\ge k_F, k\ne k_0} \frac{M_{k_0k}}{\epsilon_{k_0}-\epsilon_k}
  + 2\sum_{k< k_F} \frac{M_{k_0k}}{\epsilon_{k}-\epsilon_{k_0}}\,.
\end{multline}
 Taking the difference between the odd and even curvatures  [see  Eq.~(\ref{curvature_mb})], many of the terms cancel and we obtain Eq.~(\ref{curvature_sp}).

\subsubsection{Exact formalism}

Assuming the many-electron spectrum of the pairing model (\ref{BCS_Hamiltonian}) is known, the matrix elements of $\hat{M}_z$ are evaluated in Appendix~\ref{Sec_curvatures_exact}. The states contributing to the even curvature (\ref{curvature_even}) are all the states $|\tilde{\Psi}, k\alpha k'\alpha'\rangle_{N_e}$ with two blocked orbitals. The sum in the odd curvature (\ref{curvature_odd}) is  over all the states $|\Psi',k_0\alpha_0 k\alpha k'\alpha'\rangle_{N_e+1}$ with three blocked orbitals, one of which is  $k_0$  (the blocked orbital in $|\Omega\rangle_{N_e+1} = |\Psi,k_0\alpha_0\rangle_{N_e+1}$), and over all the states  $|\Psi'',k\alpha\rangle_{N_e+1}$ with a single blocked orbital $k\ne k_0$. 
The even curvature is then given by
\begin{equation}\label{curvature_even_exact}
 \kappa_0^{N_e} = 4 \sum_{k<k'} \sum_{\tilde{\Psi}}\frac{M_{kk'} \left|B_{kk'}[0,\tilde{\Psi}]\right|^2}{E^{N_e}_0 - E^{N_e}_{\tilde{\Psi} kk'}}\,,
\end{equation}
and the expressions (\ref{curvature_nonint_1}) and ({\ref{curvature_nonint_2}}) for the odd curvature  change to
\begin{equation}\label{curvature_odd_1_exact}
 \kappa_{\Omega}^{(1)} = 4 \sum_{k<k': k,k'\ne k_0} \sum_{\Psi'}\frac{M_{kk'} \left|B_{kk'}[\Psi,\Psi']\right|^2}{E^{N_e+1}_\Omega - E^{N_e+1}_{\Psi' k_0kk'}}
\end{equation}
and
\begin{equation}\label{curvature_odd_2_exact}
 \kappa_{\Omega}^{(2)} = 2\sum_{k\ne k_0} \sum_{\Psi''} \frac{M_{kk_0}|D_{kk_0}[\Psi,\Psi'']|^2}{E^{N_e+1}_{\Omega} - E^{N_e+1}_{\Psi'' k}}\,.
\end{equation}
Here, the denominators contain the differences between the corresponding many-body eigenenergies. The many-particle contributions to the matrix element $B_{kk'}[0,\tilde{\Psi}]$ and $D_{kk_0}[\Psi,\Psi'']$ are given, respectively, in Eqs.~(\ref{B_kk'}) and (\ref{D_kk}) of  Appendix~\ref{Sec_curvatures_exact}.
When $G=0$, they are identically zero for most of the terms, and are equal to $1$ for  the terms shown in Fig.~\ref{Fig_matrixelements}.

In the noninteracting limit, the number of many-electron states $|\Omega'\rangle_{N_e+1}$ and $|\Theta'\rangle_{N_e}$ contributing to Eqs.~(\ref{curvature_odd}) and (\ref{curvature_even})  scales as a power of $N_{\text{sp}}$, the number of single-particle orbitals in the model space.  In the presence of pairing, however, the number of  such states scales combinatorially with $N_{\text{sp}}$; any state with an allowed configuration of the blocked orbitals has generally a nonzero contribution. Furthermore, the computational effort required to evaluate each separate many-electron matrix element of $\hat{M}_z$ has a combinatorial dependence on $N_{\text{sp}}$ as well.  
We have performed exact calculations for model spaces with $N_{\text{sp}}\le 13$.

\subsubsection{BCS formalism}

 To calculate the level curvature in larger model spaces (up to $N_{\text{sp}}\sim 200$), we employ a generalized BCS approach,~\cite{Soloviev1961,Braun1997,Braun1999} in which the blocking effect in the odd states is partly accounted for. The computational time required to solve each BCS equation and the number of relevant many-electron states in the sums (\ref{curvature_odd}) and (\ref{curvature_even}) scale as powers of $N_{\text{sp}}$, enabling calculations in much larger model spaces. 
  Below we summarize  this method, and more details are presented in Appendix~\ref{Sec_curvatures_bcs}.

The BCS ground state for  even particle number $N_e$ is given by
\begin{equation}\label{BCS_even_gs}
 |\text{BCS}_e\rangle = \prod_k (u_k +v_k c^\dagger_{k1}c^\dagger_{k2})|\text{vac}\rangle\,,
\end{equation}
where $|\text{vac}\rangle$ is the vacuum state.
Here,
\begin{equation}\label{u_kv_k}
u_k^2 = \frac 12 \left(1 + \frac{\xi_k}{E_k}\right) \,,\quad
 v_k^2 = \frac 12 \left(1 - \frac{\xi_k}{E_k}\right)\,,
\end{equation}
where 
\begin{equation}\label{xi_k}
\xi_k = \epsilon_k - \mu_e - Gv_k^2\,,
\end{equation}
and $E_k$ are the quasiparticle energies 
\begin{equation}\label{E_k}
E_k = \sqrt{\xi_k^2 + \Delta_e^2}\,.
\end{equation}
The pairing gap $\Delta_e$ and the chemical potential $\mu_e$ are determined from the self-consistent BCS equations
\begin{equation}\label{BCS_eq_ev}
 \Delta_e = G\sum_k u_k v_k\quad \text{and} \quad N_e = 2\sum_k v_k^2\,.
\end{equation}

The excited states $|\Theta'\rangle_{N_e}$ contributing to Eq.~(\ref{curvature_even}) are the two-quasiparticles excitations
with the excitation energies $E^{N_e}_{\Theta'}-E^{N_e}_0$ given by the sums of the quasiparticle energies (\ref{E_k}).
The curvature of the even ground state becomes (see Appendix~\ref{Sec_curvatures_bcs})
\begin{equation}\label{curvature_even_BCS}
 \kappa_0^{N_e}\left.\vphantom{\int}\right|_{\text{BCS}}
 =- 4\sum_{k<k'} \frac{M_{kk'}(u_kv_{k'} - u_{k'}v_k)^2}{E_k+E_{k'}}\,,
\end{equation}
where $M_{kk'}$ is defined in  Eq.~(\ref{M_kk'}).
Here, the many-electron contribution  $(u_kv_{k'} - u_{k'}v_k)^2$ results in the suppression  of the terms with $\epsilon_k$ far above  or $\epsilon_{k'}$ far below the Fermi level, similarly to the noninteracting restriction $k<k_F\le  k'$. This restriction is now lifted within
the pair-scattering energy window of $\sim 2\Delta_e$ around the Fermi level.
When there is no gapped solution to the BCS equation (i.e., when $\Delta_e=0$), Eq.~(\ref{curvature_even_BCS}) reduces to the noninteracting result (\ref{curvature_nonint_even}).

The result (\ref{curvature_even_BCS}) is very similar to Belyaev formula~\cite{Belyaev1961} for the nuclear moment of inertia and to the expression for the zero-temperature spin susceptibility of a superconductor with spin-orbit scattering derived by Anderson.~\cite{Anderson1959_prl}  These formulas describe the suppression of the respective quantities caused by pairing correlations. 

For odd particle number, we perform blocked BCS calculations to account for the reduction of pairing correlations in the odd-particle number states. We consider only those final states $|\Omega\rangle_{N_e+1}$ that have the lowest energy for a given blocked orbital $k_0$; the peak heights in a tunneling spectroscopy experiment that correspond to transitions to other states are suppressed in the BCS limit $\Delta/\delta \gg 1$ (see Appendix~\ref{Sec_peak_heights}). 

The variational lowest-energy state with a blocked orbital $k_0$ is given by
\begin{equation}\label{BCS_odd}
|\text{BCS},{k_0\alpha_0}\rangle = c^\dagger_{k_0\alpha_0}\prod_{k\ne k_0} \left({u}_{k_0k} + {v}_{k_0k} c^\dagger_{k1}c^\dagger_{k2}\right) |\text{vac}\rangle\,.
\end{equation}
The corresponding pairing gap $\Delta_{k_0}$ and the chemical potential $\mu_{k_0}$ are now determined from 
\begin{equation}
{\Delta}_{k_0} = G\sum_{k\ne k_0} {u}_{k_0k}{v}_{k_0k}  \quad \text{and} \quad   N_e = 2\sum_{k\ne k_0} {v}_{k_0k}^2\,.
\end{equation}
The  parameters $u_{k_0k}$, $v_{k_0k}$, $\xi_{k_0k}$, and $E_{k_0k}$ are defined similarly to the even case for $k\ne k_0$ [Eqs.~(\ref{u_kv_k}), (\ref{xi_k}), and (\ref{E_k}) above].
 We approximate any other relevant state by a quasiparticle excitation on top of one of the states (\ref{BCS_odd}).

 Similarly to Eq.~(\ref{curvature_even_BCS}), the contribution (\ref{curvature_odd_1_exact}) to the odd curvature from pairs of states with different numbers of blocked orbitals becomes
 \begin{equation}\label{curvature_odd1_BCS}
 \kappa_\Omega^{(1)}\left.\vphantom{\int}\right|_{\text{BCS}} = - 4\sum_{k<k', \ne k_0} \frac{M_{kk'}(u_{k_0k}v_{k_0k'} - u_{k_0k'}v_{k_0k})^2}{E_{k_0k}+E_{k_0k'}}\,.
 \end{equation}
The second contribution (\ref{curvature_odd_2_exact}) reduces to (see Appendix~\ref{Sec_curvatures_bcs})
\begin{multline}\label{curvature_odd2_BCS}
 \kappa_{\Omega}^{(2)}\left.\vphantom{\int}\right|_{\text{BCS}} =
 -2\sum_{k\ne k_0} M_{k_0k} \left[ \frac{(u_{k_0k}u_{kk_0}+v_{k_0k}v_{kk_0})^2}{E^{\text{BCS}}_{k}-E^{\text{BCS}}_{k_0}}
  \right. \\ \left.
  +\frac{(u_{k_0k}v_{kk_0}-u_{kk_0}v_{k_0k})^2}{E^{\text{BCS}}_{k} + 2E_{kk_0}-E^{\text{BCS}}_{k_0}}
 \right]\,,
\end{multline}
where
\begin{equation}\label{BCS_energy}
 E^{\text{BCS}}_{k_0} = \sum_{k\ne k_0}({\xi}_{k_0k} - {E}_{k_0k} + G {v}_{k_0k}^4) + \frac{{\Delta}_{k_0}^2}{G} + {\mu}_{k_0}N_e + \epsilon_{k_0}
\end{equation}
is the BCS energy of the state (\ref{BCS_odd}). For each  blocked orbital $k\ne k_0$, two odd-particle-number doublets contribute to Eq.~(\ref{curvature_odd2_BCS}). One of them is the lowest-energy level with the blocked orbital $k$, and the other is a two-qusiparticles excitation on top of $|\text{BCS},k\alpha\rangle$ in which the two quasiparticles occupy the same orbital $k_0$, so the number of blocked orbitals does not change.
If the blocking effect in the BCS calculations were ignored, the denominators of these two contributions would be $E_k - E_{k_0}$ and $E_k+E_{k_0}$, and the entire expression would resemble  the odd-particle-number nuclear moment of inertia at zero temperature derived in Ref.~\onlinecite{Alhassid2005_prc}.

\subsection{Numerical simulations}\label{Sec_curvatures_simulations}

We carried out both exact and BCS calculations of the level curvature.  For each calculation, we generate an ensemble of 1000 $2N_{\text{rmt}}\times 2N_{\text{rmt}}$ GSE random matrices with $N_{\text{rmt}}=201$ degenerate eigenvalues each. The matrices are diagonalized by  symplectic transformations using the phase convention (\ref{phase_convention}) for the eigenvectors. For each matrix, we form the model space of $N_{\text{sp}} < N_{\text{rmt}}$ single-particle Kramers doublets by taking $N_{\text{sp}}$ levels in the middle of the spectrum and unfolding~\cite{Bohigas1984} the eigenvalues. We use $N_{\text{sp}}=13$ in the exact numerical calculations and $N_{\text{sp}}=121$ in the  BCS calculations.  Assuming the orbital contribution to the magnetization is negligible, we use the eigenvectors to calculate the single-particle matrix elements of $\hat{M}_z = 2\mu_B\hat{S}_z$ in the basis of $2N_{\text{sp}}$ states forming the chosen doublets. 
We perform the many-body calculations for half-filling (i.e., $N_e=N_{\text{sp}}-1$ for odd $N_{\text{sp}}$). In the exact calculations, we use the Lanczos algorithm~\cite{Lehoucq1998} to calculate the relevant many-electron eigenfunctions and eigenenergies. 

We select the odd eigenstates $|\Omega\rangle_{N_e+1}$  based on the heights of the
differential-conductance peaks, which are calculated as discussed in Appendix~\ref{Sec_peak_heights}.  We select only those states whose peak height is at least 10\% of the average peak height for an allowed transition in the noninteracting limit.
In that limit,  a final state $|\Omega\rangle_{N_e+1}$
has the lowest energy among all the states with the same single blocked orbital $k_0$ ($k_0 \geq k_F$). Similarly, in the BCS limit $\Delta/\delta \gg 1$, the peak heights are nonzero only for the lowest-energy states for a given $k_0$, although  now $k_0$ may also lie below $k_F$ (the peak height is suppressed when $\epsilon_{k_F}-\epsilon_{k_0} \gg \Delta$). 
In the BCS calculations, we consider only such lowest-energy states, while in the exact calculations we consider general states with a single blocked orbital.

We sort all resolvable odd-particle-number doublets according to their energies. We refer to the curvature for the transition $|0\rangle_{N_e} \rightarrow |\Omega\rangle_{N_e+1}$ to a state in the $n$-th doublet as to the curvature of the $n$-th differential-conductance peak. In this convention, the first  peak usually corresponds to the tunneling into the odd ground state.

We characterize each curvature distribution  by its median value $\kappa_{\text{med}}$ and midspread (i.e., the width of the middle part of the distribution with 50\% of the total area) $d_{\kappa}$, which are robust measures for a distribution with possibly long tails.

To eliminate the nonphysical dependence on the random-matrix size, we express  our results in units of $d_0$, the  midspread of the  single-particle level curvature distribution. In the GSE limit, this $d_0$ is related to the $g$-factor statistics by~\footnote{
The relation between the single-particle level midspread $d_0$ and variance {$\langle|\kappa^2|\rangle$} can be found from the known distribution of the single-particle level curvature~\cite{Fyodorov1995,vonOppen1995} resulting in {$d_0 \approx 1.13\sqrt{\langle|\kappa^2|\rangle}$}.
 The curvature variance for a single-particle level $k_0$ can be related to {$\langle g^2\rangle$} using the formalism of Ref.~{\onlinecite{Mucciolo2006}}.  We note, however, that in  Ref.~{\onlinecite{Mucciolo2006}}, the RMT average {$\left\langle\sum_{k \ne k_0}\sum_{k' \ne k_0, k}{1}/{[(\epsilon_{k_0}-\epsilon_{k})(\epsilon_{k_0}-\epsilon_{k'})]}\right\rangle$} was taken to be zero (leading to Eq.~(15) of that reference) since the average over the level $k_0$ was taken over the entire spectrum of the matrix.  We find numerically that this RMT average is  {$-(3/4)\left\langle \sum_{k \ne k_0} 1/[(\epsilon_{k_0}- \epsilon_{k})^2] \right\rangle$} when the level $k_0$ is taken in the middle of the spectrum of a large matrix. As a result, the right-hand side of Eq.~(20) of Ref.~{\onlinecite{Mucciolo2006}} is reduced by a factor of 2 (also note that we use a different definition of the parameter $\delta$).
}
\begin{equation}
 d_0 \approx 0.68 \frac{\mu_B^2}{\delta} \langle g^2 \rangle\,.
\end{equation}
 This relation also holds when there is an orbital contribution to the magnetization.

When the single-particle levels follow RMT statistics, the average matrix elements of the spin operator do not decay with the energy separation between two single-particle levels, and  the contribution from single-particle levels far from $k_F$ may be important.
Using BCS calculations in the presence of pairing correlations, we found that 
the median values converge slowly with increasing values of $N_{\text{sp}}$, and we were unable to determine whether convergence is reached even at  $N_{\text{sp}} \sim 200$.  In contrast, the convergence of the  midspreads is clearly reached at $N_{\text{sp}}\sim 100$, and the results  at $N_{\text{sp}}\sim 10$ are already quite close to their asymptotic values.  We, therefore, consider the median values  only as  qualitatively correct in both the exact and the BCS calculations, while the results for the midspreads are expected to be reliable even in the relatively small single-particle model space used in the exact calculations.

\subsection{Results and discussion}\label{Sec_curvatures_results}

In the following we present results for the level curvature statistics.

\subsubsection{First differential-conductance peak}

\begin{figure}
\includegraphics[clip,width=0.45\textwidth]{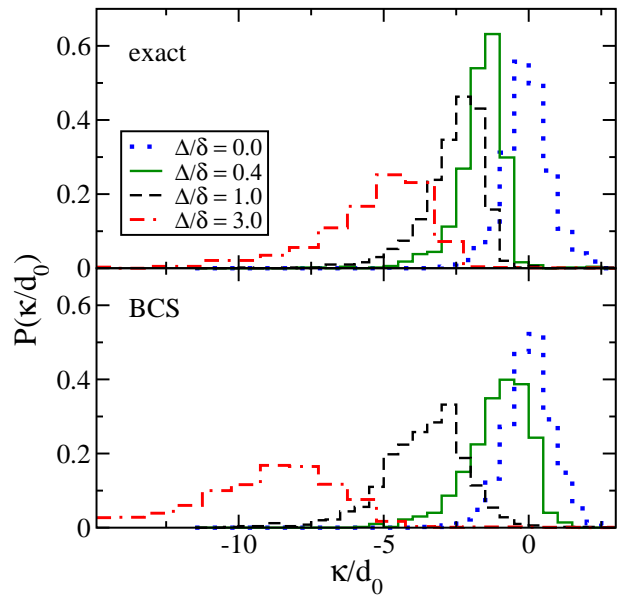}
\caption{The level curvature distribution $P(\kappa/d_0)$ of the first differential-conductance peak  for $\Delta/\delta=0$ (dotted lines), $\Delta/\delta  = 0.4$ (solid lines), $\Delta/\delta = 1$  (dashed lines), and $\Delta/\delta=3$ (dash-dotted lines). The curvature is expressed in units of the single-particle midspread $d_0$ (the width of the middle 50\% of the distribution). The top panel describes the results of the exact diagonalization method and the bottom panel  describes the BCS results.}\label{Fig_distgaps}
\end{figure}

\begin{figure}
\includegraphics[clip,width=0.45\textwidth]{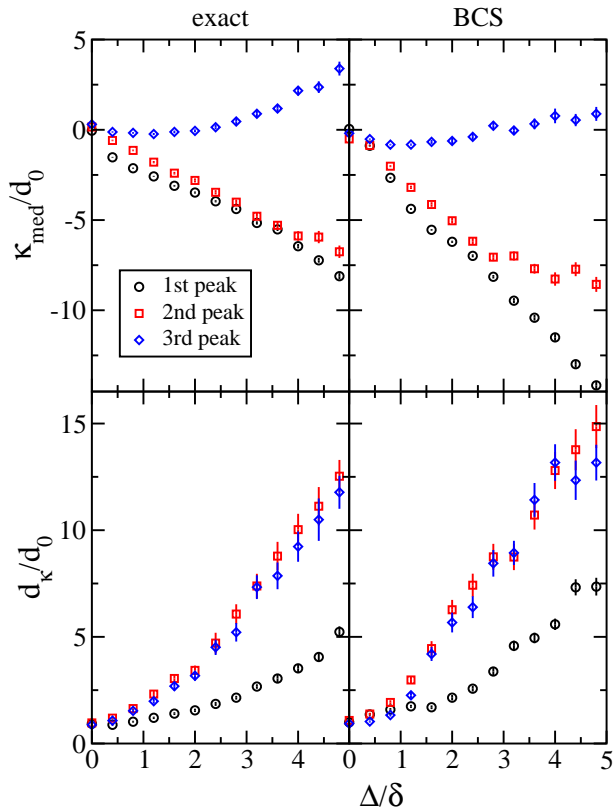}
\caption{The median  $\kappa_{\text{med}}$ (top panels) and midspread $d_{\kappa}$ (bottom panels) of the level curvature distribution versus  $\Delta/\delta$. The results are shown in units of  $d_0$ for the first (circles), second (squares), and third (diamonds) differential-conductance peaks, using the exact (left column) and BCS (right column) formalisms. The statistical errors~\cite{Stuart1987} are shown by vertical bars. The results at different values of $\Delta/\delta$ are uncorrelated, i.e., they are obtained using different ensembles of random matrices.}\label{Fig_gapdep}
\end{figure}

The level curvature distribution for the first differential-conductance peak is shown in Fig.~\ref{Fig_distgaps} for four values of  $\Delta/\delta$ using exact diagonalization (top panel) and the BCS approximation (bottom panel).   The median and midspread  are shown by the open circles in Fig.~\ref{Fig_gapdep}  as a function of $\Delta/\delta$. 

We observe the following qualitative features in both exact and BCS results: (i) The level curvature distribution, which is symmetric around zero at $\Delta/\delta=0$, shifts to negative values in the presence of pairing correlations and becomes asymmetric with an extended tail at negative values.
(ii) The modulus of the median and
the dispersion of the distribution increase monotonically with  $\Delta/\delta$.

These observations are explained by the BCS approximation  
 for the transition  $|0\rangle_{N_e} \rightarrow |0\rangle_{N_e+1}$ as follows. The excitations that contribute to the curvature of the even ground state (\ref{curvature_even_BCS}) are two-quasiparticle excitations, whose energies are at least $2\Delta_e$. When pairing is sufficiently strong, this suppresses the positive ($-\kappa_0^{N_e}>0$) contribution to the observed curvature $\kappa_0^{N_e+1}-\kappa_0^{N_e}$ and, in particular, reduces the right tail of the distribution. For odd particle number, there are  excitations that involve the change of the blocked orbital without changing the number of Cooper pairs [see Eq.~(\ref{curvature_odd2_BCS}) and Figs.~\ref{Fig_matrixelements}(c) and \ref{Fig_matrixelements}(d)].
 In the limit $\Delta/\delta \gg 1$, the lowest excitation energy for the odd grain 
can then be estimated as the difference between two quasiparticle energies $E_{k_F+1}-E_{k_F} \approx \sqrt{\delta^2+\Delta^2}-\Delta \approx \delta^2/(2\Delta)$, which is much smaller than the noninteracting value $\delta$. Therefore, the negative contribution to the level curvature ($\kappa_0^{N_e+1}<0$) is enhanced by pairing correlations.  When mesoscopic fluctuations are taken into account, the lowest excitation energy fluctuates and can assume even smaller values, thus extending the left tail of the distribution.

In the fluctuation-dominated regime $\Delta/\delta \lesssim 1$,  pairing correlations are noticeably weaker in an odd grain than in an even grain because of the blocking effect. 
This can be observed in the exact results, which are more reliable than the BCS results in this regime. For $\Delta/\delta=0.4$,  pairing correlations are sufficiently strong in the even states $|0\rangle_{N_e}$ so as to suppress the right tail of the distribution. However,  in the odd states $|0\rangle_{N_e+1}$ they are not sufficiently strong to extend the left tail of the distribution or to increase the total dispersion (see the top panel of Fig.~\ref{Fig_distgaps}). 

 It is remarkable that the asymmetry effect is strong even in the fluctuation-dominated regime $\Delta/\delta <1$, where the probability to observe a positive level curvature is small (see the exact results for $\Delta/\delta=0.4$ in Fig.~\ref{Fig_distgaps}). We conclude that the level curvature of the first peak is a sensitive probe to detect pairing correlations in the fluctuation-dominated regime. This probe is practical even when only a few data points are available. 
  
For a single grain, level curvature statistics can be collected by changing the gate voltage and thus varying the number of  electrons  in the grain. An alternative way to observe the effect is to fix some orbital $k_0$ and study the curvature for the transition $|0\rangle_{N_e} \rightarrow |\Psi,k_0\alpha_0\rangle_{N_e+1}$ as a function of $N_e$, where $|\Psi,k_0\alpha_0\rangle_{N_e+1}$ is the lowest-energy state for given $k_0$ and $N_e$. When the Fermi level is tuned between a value far below $k_0$ and  $k_0$  so $|\Psi,k_0\alpha_0\rangle_{N_e+1}$ becomes the odd ground state, the total change in the curvature should be negative. 
  For each value of $N_e$, the  differential-conductance peak with the blocked orbital $k_0$ can be identified by its $g$-factor $g_{k_0}$, which is independent of the position of the Fermi level (see  Sec.~\ref{Sec_gfactors}).

\subsubsection{Higher differential-conductance peaks}

We next discuss the level curvature statistics for the second and third differential-conductance peaks. The median values and midspreads are shown in Fig.~\ref{Fig_gapdep} as a function of $\Delta/\delta$ by squares (second peak) and diamonds (third peak).

We observe that the level curvature dispersions for the second and third peaks behave alike as a function of $\Delta/\delta$ and are larger than the level curvature dispersions for the first peak. However,  the median values for the second and third peaks behave very differently from each other. To understand these results, 
we consider the odd excited states $|2\rangle_{N_e+1}$ and $|3\rangle_{N_e+1}$ that correspond to the second and third peaks with energies  $E^{N_e+1}_{(2)}$ and $E^{N_e+1}_{(3)}$,  respectively, and denote by $k_0$  the ground-state blocked orbital. When pairing correlations are weak, the blocked orbitals in the states $|2\rangle_{N_e+1}$ and $|3\rangle_{N_e+1}$ are usually  $k_0+1$ and $k_0+2$, while the state with the blocked orbital $k_0-1$ is unresolved. 
However, when pairing is sufficiently strong, the blocked orbitals in the states  $|2\rangle_{N_e+1}$ and $|3\rangle_{N_e+1}$ tend to be $k_0-1$ and $k_0+1$.  
In the limit $\Delta/\delta \gg 1$ and for an equally spaced single-particle spectrum, the energies $E^{N_e+1}_{(2)}$ and $E^{N_e+1}_{(3)}$ are equal since the  quasiparticle energies $E_{k_0-1}$ and $E_{k_0+1}$ are both $\sqrt{\delta^2 + \Delta^2}$. 
In a realistic RMT-like spectrum and for finite but sufficiently large $\Delta/\delta$, these energies are not equal, but tend to be closer to each other than to other eigenenergies. 
Therefore, the term in Eq.~(\ref{curvature_odd})  that contains the difference between $E^{N_e+1}_{(2)}$ and $E^{N_e+1}_{(3)}$ in the denominator is often the dominant term. 
Since such terms in the perturbation theory expressions have equal amplitudes but opposite signs for the second and third peaks, the dispersions of the resulting distributions behave similarly, while typical values of the corresponding curvatures are very different. 

\begin{figure}
 \includegraphics[clip,width=0.45\textwidth]{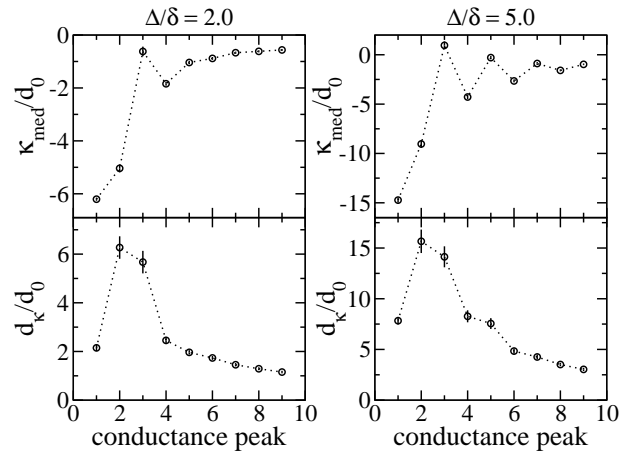}
\caption{The median level curvature $\kappa_{\text{med}}$ (top panels) and midspread $d_{\kappa}$ (bottom panels) in units of  $d_0$ as a function of the differential-conductance-peak number. The results are obtained using the BCS formalism for $\Delta/\delta=2$ (left column) and $\Delta/\delta=5$ (right column). }\label{Fig_peakdep}
\end{figure}

Similar effect can be observed in higher differential-conductance peaks. In Fig.~\ref{Fig_peakdep}, we show the medians and midspreads of the level curvature distributions as a function of the peak number for the first nine resolved peaks for $\Delta/\delta=2$ (left panel) and $\Delta/\delta=5$ (right panel). 
We observe similar values for the level curvature dispersions of the $2n$-th and $(2n+1)$-th peaks. However, the level curvature median has a negative contribution for the $2n$-th peak and a positive contribution for the $(2n+1)$-th peaks. 
This latter effect is enhanced at larger values of $\Delta/\delta$, leading to odd-even staggering in $\kappa_{\rm med}$ vs.~conductance peak number.  As the peak number increases, both the midspreads and medians  converge to their noninteracting (single-particle) values.
 
\section{Conclusion}\label{Sec_conclusions}

We have studied the effect of pairing correlations on the 
magnetic-field response of discrete energy levels in a metallic nanoparticle with spin-orbit scattering.  In particular, we investigated the $g$-factor and zero-field level curvature statistics, which parametrize, respectively, the first- and second-order corrections (in the magnetic field) to the energies measured in single-electron-tunneling spectroscopy experiments. 

We have shown that $g$-factors are not affected by pairing correlations and reduce to the $g$-factors of the single-particle orbitals.  This conclusion follows from the blocking effect of pairing correlations and from considerations of time-reversal symmetry.  It is independent of the strength of spin-orbit scattering and of the relative effects of orbital and spin magnetisms.  Thus, we can use  $g$-factor measurements to probe the importance of electron-electron correlations beyond pairing interactions. 

In contrast, level curvatures are highly sensitive to pairing correlations. This can be understood qualitatively by the dependence of the level curvature on the density of states, which in turn is modified by pairing correlations. In particular, the level curvature distribution for the first differential-conductance peak,  which in the absence of pairing correlations is symmetric around zero, shifts almost entirely to negative values even in the fluctuation-dominated regime $\Delta/\delta <1$. In this regime, the pairing-induced excitation gap cannot be resolved in a conventional spectroscopy experiment, while the change in the level curvature can still be observed and can therefore be used to probe pairing correlations. We have demonstrated these results by using both exact diagonalization and a BCS approximation. The latter approach can be applied in a much larger single-particle model space.  

The exact and BCS formalisms discussed here can be applied in the more general case of arbitrary spin-orbit scattering  and in the presence of orbital magnetism.  
Another possible application of the techniques developed here is in the calculation of the low-temperature spin susceptibility of a superconducting nanoparticle with spin-orbit scattering. The non-divergent contributions to the spin susceptibility in the limit $k_BT\ll \delta$ ($k_B$ is the Boltzmann constant and $T$ is  temperature) resemble the perturbative expressions for the level curvature [Eqs.~(\ref{curvature_odd}) and (\ref{curvature_even})]. For odd particle number, there is also a divergent Curie-like contribution $\sim \langle\hat{S}_z^2\rangle/T$, which is suppressed for strong spin-orbit scattering.  The odd-even effect in the spin susceptibility of normal-metal particles with spin-orbit scattering was  studied in Ref.~\onlinecite{Sone1977}, where  it was shown that  both the even and odd susceptibilities reduce to the high-temperature Pauli susceptibility in the limit of strong spin-orbit scattering. This corresponds here to the zero average curvature $\kappa_0^{N_e+1}-\kappa_0^{N_e}$ in the 
noninteracting limit. The spin 
susceptibility of superconducting particles with spin-orbit scattering was studied in Ref.~\onlinecite{Shiba1976}, but without considering the odd-even effect.

\begin{acknowledgments}
We thank  L.I. Glazman for useful discussions.
This work was supported in part by the U.S. DOE grant No. DE-FG02-91ER40608. Computational cycles were provided by the facilities of the Yale University Faculty of Arts and Sciences High Performance Computing Center.
\end{acknowledgments}

\appendix


\section{Many-particle matrix elements of $\hat{M}_z$ in the exact formalism}\label{Sec_curvatures_exact}

Here we derive the expressions for the many-particle matrix elements of $\hat M_z$ between exact many-particle eigenstates of the even and odd grain.

\subsection{Even particle number}

We denote by $\overline{\alpha}$ the time-reversed state of $\alpha$ (e.g., $\overline{1}=2$) and by  $\vec{m'}\cup\{k\}$ the set of orbitals containing the set $\vec{m'}$ and one extra orbital $k$ that is not in $\vec{m'}$.
Using Eqs.~(\ref{eigenstate_even_ground}), (\ref{one_body_operator}), and  (\ref{Mz_sp_timereversal}),  and the relation 
\begin{equation}\label{TRS_matrelement}
M^z_{k\alpha,k'\alpha'} = (-1)^{\alpha+\alpha'+1}\left(M^z_{k\overline{\alpha},k'\overline{\alpha}'}\right)^*\,,
\end{equation}
which is valid for the phase convention (\ref{phase_convention}), we find
\begin{multline}
\hat{M}_z|0\rangle_{N_e} = \sum_{k<k'}\sum_{\alpha\alpha'} (-1)^{\alpha'-1} M^z_{k\alpha,k'{\alpha}'}\\
 \times\sum_{\vec{m'}: k,k'\notin \vec{m'}} \left(C^0_{\vec{m'}\cup\{k'\}} - C^0_{\vec{m'}\cup\{k\}}\right)|\vec{m}',k\alpha k'\overline{\alpha}'\rangle_{N_e}\,.
\end{multline}
 This sum over Slater-determinant states is over the positions of two blocked orbitals and configurations of the remaining $N_e/2-1$ electron pairs. 
Therefore, a state $|\Theta'\rangle_{N_e}$ with non-zero matrix element $\langle\Theta'|\hat{M}_z|0\rangle_{N_e}$ must have two unpaired electrons. The matrix element between $|0\rangle_{N_e}$ and another fully-paired state is identically zero because of the time-reversal symmetry. 
The matrix element of $\hat{M}_z$ between the ground state and any state with two blocked orbitals
\begin{equation}
|\tilde{\Psi},k\alpha k'\alpha' \rangle_{N_e} = \sum_{\vec{m'}: k,k'\notin \vec{m'}}C^{\tilde{\Psi}}_{\vec{m}'} |\vec{m}',k\alpha k'{\alpha}'\rangle_{N_e}
\end{equation}
is thus given by
\begin{equation}\label{matr_element_2blocked}
\langle\Psi',k\alpha k'\alpha' |\hat{M}_z|0\rangle_{N_e}  = (-1)^{\alpha'}M^z_{k\alpha,k'\overline{\alpha}'} B_{kk'}[0,\tilde{\Psi}]\,,
\end{equation}
where
\begin{equation}\label{B_kk'}
B_{kk'}[0,\tilde{\Psi}] =  \sum_{\vec{m'}: k,k'\notin \vec{m'}} C^{\tilde{\Psi}}_{\vec{m}'}\left(C^0_{\vec{m'}\cup\{k'\}} - C^0_{\vec{m'}\cup\{k\}}\right)^*
\end{equation}
describes the interaction contribution to the matrix element.

\subsection{Odd particle number}

For a given blocked orbital $k_0$ in an odd-grain state (\ref{odd_state}), we divide the sum in Eq.~(\ref{one_body_operator}) into three contributions
\begin{equation}\label{M_z_three_groups}
 \hat{M}_z = \hat{M}_z^{(1)} + \hat{M}_z^{(2)} + \hat{M}_z^{(3)}\,.
\end{equation}
Here $\hat{M}_z^{(1)}$ consists of terms  that do not contain  $k_0$,   $\hat{M}_z^{(2)}$ consists of terms that produce the linear correction (i.e., with $c^\dagger_{k_0\alpha}c_{k_0\alpha'}$), and $\hat{M}_z^{(3)}$ consists of the remaining terms.
By analogy with the even case, the first contribution results in nonzero matrix elements 
between $|\Psi,k_0\alpha_0\rangle_{N_e+1}$  and the states with three blocked orbitals (with the number of Cooper pairs reduced by one)
\begin{multline}
 \langle\Psi',k\alpha k'\alpha'k_0\alpha_0 |\hat{M}_z|\Psi,k_0\alpha_0\rangle_{N_e+1} \\ = (-1)^{\ldots} M^z_{k\alpha,k'\overline{\alpha}'}
  B_{kk'}[\Psi,\Psi']\,.
\end{multline}
Here $B_{kk'}[\Psi,\Psi']$ is defined as in Eq.~(\ref{B_kk'}) but with $C^0$'s substituted by $C^\Psi$'s of Eq.~(\ref{odd_state}).
The phase of the matrix element depends on the relative positions of the three blocked orbitals and is not important.
The second group of terms in Eq.~(\ref{M_z_three_groups}) does not contribute to the level curvature,  while the
 third contribution is
\begin{multline}
\hat{M}_z^{(3)} |\Psi,k_0\alpha_0\rangle_{N_e+1} = \sum_{k\alpha, k\ne k_0} M^z_{k\alpha,k_0\alpha_0} \\
 \times \left(\sum_{\vec{m}: k,k_0 \notin \vec{m}} C^\Psi_{\vec{m}} |\vec{m},k\alpha\rangle_{N_e+1}  \right. \\
\left. + \sum_{\vec{m}: k\in \vec{m}, k_0\notin \vec{m}} C^{\Psi}_{\vec{m}}|\vec{m}\backslash\{k\}\cup\{k_0\}, k\alpha\rangle_{N_e+1}
   \right)\,.
\end{multline}
Here the set $\vec{m}\backslash\{k\}\cup\{k_0\}$ is obtained from $\vec{m}$ by replacing $k$ with $k_0$. Therefore, the matrix element between two states with single but different blocked orbitals is given by
\begin{equation}
 \langle \Psi'',k\alpha |\hat{M}_z|\Psi,k_0\alpha_0\rangle_{N_e+1} = M^z_{k\alpha,k_0\alpha_0} D_{k_0k}[\Psi,\Psi'']\,,
 \end{equation}
 where
 \begin{equation}\label{D_kk}
 D_{k_0k}[\Psi,\Psi''] = \sum_{\vec{m}:k_0\notin \vec{m}} C^{\Psi''}_{\vec{m}''}C^{\Psi}_{\vec{m}}
\end{equation}
and
\begin{equation}
 \vec{m}'' = \left\{ \begin{array}{l}
                     \vec{m} \quad \text{if}\quad k\notin \vec{m} \\
		     \vec{m}\backslash \{k\}\cup\{k_0\} \quad \text{if} \quad k \in \vec{m}\,.
                    \end{array}\right.
\end{equation}


\section{Details of the BCS formalism}\label{Sec_curvatures_bcs}
\subsection{Even curvature}

The state $|\text{BCS}_e\rangle$ defined in Eq.~(\ref{BCS_even_gs}) is the vacuum with respect to the quasiparticle operators 
\begin{equation}\label{quasiparticle_operators}
 \left\{
 \begin{array}{l}
  \beta_{k1} = u_{k}c_{k1} - v_{k} c^\dagger_{k2}\,,\\
  \beta_{k2} = u_{k} c_{k2} + v_{k} c^\dagger_{k1}\,.
 \end{array}
 \right.
\end{equation}
Using the inreverse transformation, the identity $\beta_{k\alpha}|\text{BCS}_e\rangle =0$, and Eq.~(\ref{TRS_matrelement}), we find
\begin{multline}
 \hat{M}_z|\text{BCS}_e\rangle  = \sum_{k < k'}(u_{k} v_{k'} - u_{k'}v_{k}) \\
 \times \left[M^z_{k1,k'1} \beta^\dagger_{k1}\beta^\dagger_{k'2}
 + \left(M^z_{k1,k'1}\right)^*\beta^\dagger_{k2}\beta^\dagger_{k'1}  -  M^z_{k1,k'2}\beta^\dagger_{k1}\beta^\dagger_{k'1}
 \right. \\
 \left.
 + \left(M^z_{k1,k'2}\right)^*\beta^\dagger_{k2}\beta^\dagger_{k'2}\right] |\text{BCS}_e\rangle \,,
\end{multline}
which leads to Eq.~(\ref{curvature_even_BCS}).

\subsection{Odd curvature}

For the odd state $|\text{BCS},k_0\alpha_0\rangle$ defined in Eq.~(\ref{BCS_odd}), we split $\hat{M}_z$ into three components according to Eq.~(\ref{M_z_three_groups}). The effect of $\hat{M}_z^{(1)}$ is similar to the effect of $\hat{M}_z$ for even particle number, resulting in the contribution~(\ref{curvature_odd1_BCS}). The second part --  $\hat{M}_z^{(2)}$ -- does not contribute to the level curvature. For the remaining part, we find [using Eqs.~(\ref{TRS_matrelement}) and (\ref{BCS_odd})]
\begin{multline}
\hat{M}_z^{(3)}|\text{BCS},k_01\rangle \\
=  \sum_{k\alpha,k\ne k_0} M^z_{k\alpha,k_01} c^\dagger_{k\alpha}  (u_{k_0k} + v_{k_0k}c^\dagger_{k_01}c^\dagger_{k_02}) \\
\times \prod_{k'\ne k,k_0}(u_{k_0k'}+v_{k_0k'}c^\dagger_{k_01}c^\dagger_{k_02})|\text{vac}\rangle\,.
\end{multline}
Therefore $\hat{M}_z^{(3)}$ has nonzero matrix elements between the states with different singly blocked orbitals.
To reduce the computational effort, we keep only those contributions that would be nonzero if the blocking effect were ignored, i.e., if all  $u_{k_0k}$'s and $v_{k_0k}$'s were equal to $u_k$'s and $v_k$'s. For a given blocked orbital $k\ne k_0$, there are two states with such matrix elements. 
The first is the lowest-energy doublet $|\text{BCS},k\alpha\rangle$ with  the variational energy $E^{\text{BCS}}_k$ defined in Eq.~(\ref{BCS_energy}).
The corresponding  matrix element is
\begin{multline}
 \langle \text{BCS},k\alpha |\hat{M}_z|\text{BCS},k_01\rangle = M^z_{k\alpha,k_01}\\ \times (u_{k_0k}u_{kk_0} +v_{k_0k}v_{kk_0})
 \prod_{k'\ne k,k_0}(u_{k_0k'}u_{kk'}+v_{k_0k'}v_{kk'})\,.
\end{multline}
The second state is the two-quasiparticle excitation on top of $|\text{BCS},k\alpha\rangle$ with both quasiparticles  on the same orbital $k_0$  (so $k_0$ is not a blocked orbital)
\begin{multline}
|\widetilde{\text{BCS}},k\alpha\rangle
=c^\dagger_{k\alpha}(-v_{kk_0}+u_{kk_0}c^\dagger_{k_01}c^\dagger_{k_02})\\ \times\prod_{k'\ne k,k_0} (u_{kk'}+v_{kk'}c^\dagger_{k'1}c^\dagger_{k'2})|\text{vac}\rangle\,.
\end{multline}
The energy of such a state is given by $E^{\text{BCS}}_k+2E_{kk_0}$, and the corresponding matrix element is
\begin{multline}
\langle \widetilde{\text{BCS}},k\alpha |\hat{M}_z|\text{BCS},k_01\rangle = M^z_{k\alpha,k_01}\\ \times ( u_{k_0k}v_{kk_0} -u_{kk_0}v_{k_0k})
 \prod_{k'\ne k,k_0}(u_{k_0k'}u_{kk'}+v_{k_0k'}v_{kk'})\,.
\end{multline}

Assuming that
\begin{equation}
 \prod_{k'\ne k,k_0}(u_{k_0k'}u_{kk'}+v_{k_0k'}v_{kk'}) \approx 1\,,
\end{equation}
which would be an identity if the blocking effect were ignored, we obtain the second contribution 
(\ref{curvature_odd2_BCS}) to the curvature of the odd state.

\section{Differential-conductance peak heights}\label{Sec_peak_heights}

Assuming a point contact between the grain and the electrode, the height of the differential-conductance peak for the transition $|0\rangle_{N_e}\rightarrow |\Omega\rangle_{N_e+1}$ in a tunneling-spectroscopy experiment is proportional to~\cite{vonDelft2001}
\begin{equation}
 w_{|0\rangle\rightarrow |\Omega\rangle} = |\langle \Omega |\psi^\dagger_{\uparrow}({\bf r})|0\rangle_{N_e}|^2 + |\langle \Omega |\psi^\dagger_{\downarrow}({\bf r})|0\rangle_{N_e}|^2\,,
\end{equation}
where $\bf r$ is the position of the contact. 
The field operators are given by a symplectic transformation [using the phase convention (\ref{phase_convention})]
\begin{eqnarray}
  \psi^\dagger_{\uparrow}({\bf r}) &=& \sum_k \left[p^*_{k}({\bf r})c^\dagger_{k1} + q^*_{k}({\bf r})c^\dagger_{k2}\right]\,, \\
 \psi^\dagger_{\downarrow}({\bf r}) &=& \sum_k \left[-q_{k}({\bf r})c^\dagger_{k1} + p_{k}({\bf r})c^\dagger_{k2}\right]\,.
\end{eqnarray}
Here $[p_k({\bf r})\; -q_k^*({\bf r})]^T$ and $[q_k({\bf r})\; p^*_k({\bf r})]^T$ are the single-particle spinor eigenfunctions written in a good-spin basis.  

Using Eqs. (\ref{eigenstate_even_ground}) and  (\ref{odd_state}),  we find
\begin{equation}\label{peak_height}
  w_{|0\rangle\rightarrow |\Omega\rangle} = w_{k_0}({\bf r})\left(\sum_{\vec{m}: k_0\notin \vec{m}}C^\Psi_{\vec{m}}C^0_{\vec{m}}\right)^2\,,
\end{equation}
where
\begin{equation}\label{peak_height_sp}
w_{k_0}(r)= |p_{k_0}({\bf r})|^2 + |q_{k_0}({\bf r})|^2\,.
\end{equation}
The peak height is now conveniently expressed as the product of a single-particle and many-body contributions.
The single-particle contribution $w_{k_0}({\bf r})$ is simply the probability density for an electron in the orbital $k_0$. 
The many-body contribution is generally nonzero for any eigenstate with a single blocked orbital. In the noninteracting limit, it is nonzero only for states $|\Omega\rangle_{N_e+1}$ shown in Fig.~\ref{Fig_states}(c), for which it is equal to 1 and the total peak height  reduces to  $w_{k_0}({\bf r})$.

In the BCS formalism, the even ground state is given by Eq.~(\ref{BCS_even_gs}),
and any odd state with a single blocked orbital $k_0$ (i.e., not only the lowest-energy state) can be written as
\begin{equation}\label{BCS_odd_k0_general}
  |\Omega\rangle_{N_e+1} = c^\dagger_{k_0\alpha_0}\prod_{k\ne k_0} \left(\tilde{u}_k + \tilde{v}_k c^\dagger_{k1}c^\dagger_{k2}\right) |\text{vac}\rangle\,.
\end{equation}
For these states, we obtain
\begin{equation}\label{peak_height_BCS}
 w_{|0\rangle\rightarrow |\Omega\rangle} = w_{k_0}({\bf r})\, u_{k_0}^2 \prod_{k\ne k_0} (u_k \tilde{u}_k + v_k\tilde{v}_k)^2\,.
\end{equation}

In the limit $\Delta/\delta \gg 1$, the blocking effect is negligible and all the odd states are the quasiparticle excitations built on top of  $|\text{BCS}_e\rangle$. For the one-quasiparticle excitation $\beta^\dagger_{k_0\alpha_0}|\text{BCS}_e\rangle$, $\tilde{u}_k = u_k$ and $\tilde{v}_k = v_k$, and the peak height reduces to
\begin{equation}\label{peak_height_BCS_limit}
 w_{|0\rangle\rightarrow |\Omega\rangle} = w_{k_0}({\bf r})\, u_{k_0}^2\,.
\end{equation}
Any other odd state (\ref{BCS_odd_k0_general}) with the same $k_0$ can be written as $\beta^\dagger_{k_0\alpha_0}\beta^\dagger_{k'1}\beta^\dagger_{k'2} \ldots|\text{BCS}_e\rangle$ for some $k'\ne k_0$. The product $\beta^\dagger_{k'1}\beta^\dagger_{k'2}$ results in 
$\tilde{u}_{k'} = -v_{k'}$ and $\tilde{v}_{k'} = u_{k'}$, which yields $w_{|0\rangle\rightarrow |\Omega\rangle} =0$.

We conclude that,  in the limit $\Delta/\delta \gg 1$,  a final state  must have the lowest energy among the states with the same blocked orbital. The interaction contribution to the peak height (\ref{peak_height_BCS_limit}) is given by $u_{k_0}^2$; therefore, the peak height for such a state is suppressed when the blocked orbital is far below the Fermi level and reduces to the noninteracting value when it is far above the Fermi level. 

In the RMT framework, the single-particle contribution (\ref{peak_height_sp}) is derived from a component of the spinor eigenvector representing the state $|k_0\alpha_0\rangle_{1}$. It is not important which component is chosen. In any of the Gaussian ensembles, the probability for $w_{k_0}({\bf r})$ to be small is exponentially suppressed.

\bibliography{literature_spinorbit}

\end{document}